# Silicone phantoms fabricated with multi-material extrusion 3D printing technology mimicking imaging properties of soft tissues in CT

Sepideh Hatamikia[1,2 *], Laszlo Jaksa[1], Gernot Kronreif[1], Wolfgang Birkfellner[3], Joachim Kettenbach[4, 5], Martin Buschmann[5], Andrea Lorenz[1]

[1]Austrian Center for Medical Innovation and Technology, Wiener Neustadt, Austria,

[2]Research Center for Medical Image Analysis and Artificial Intelligence (MIAAI), Department of Medicine, Danube Private University, Krems, Austria

[3]Center for Medical Physics and Biomedical Engineering, Medical University of Vienna, Vienna, Austria,

[4]Institute of Diagnostic, Interventional Radiology and Nuclear Medicine, Landesklinikum Wiener Neustadt, Wiener Neustadt, Austria,

[5]Center for Radiology, Faculty of Medicine and Dentistry, Danube Private University, 3500 Krems, Austria

[6]Department of Radiation Oncology, Medical University of Vienna/AKH Wien, Vienna, Austria

## Abstract

Recently, 3D printing has been widely used to fabricate medical imaging phantoms. So far, various rigid 3D printable materials have been investigated for their radiological properties and efficiency in imaging phantom fabrication. However, flexible, soft tissue materials are also needed for imaging phantoms, which are used in various scenarios, such as anatomical deformations to improve dynamic treatments and various needle-based surgeries and training. Recently, various silicone additive manufacturing technologies have been used to produce anatomical models based on extrusion techniques that allow the fabrication of soft tissue materials. To date, there is no systematic study in the literature investigating the radiological properties of silicone rubber materials/fluids for imaging phantoms fabricated directly by extrusion using 3D printing techniques. The aim of this study was to investigate the radiological properties of 3D printed phantoms made of silicone in CT imaging. The radiodensity as described as Hounsfield Units (HU) of several test phantoms composed of three different silicone printing materials were evaluated by changing the infill density to adjust their radiological properties. A comparison of HU values with a Gammex Tissue Characterization Phantom was performed. A scaled down anatomical model derived from an abdominal CT was also fabricated and the resulting HU values were evaluated. A reproducibility analysis was also performed by creating several replicas for specific infill densities. For the three different silicone materials, a spectrum ranging from -639 to +780 HU was obtained on CT at a scan setting of 120 kVp. A good agreement was observed between the HU target values in abdominal CT and the HU values of the 3D-printed anatomical phantom in all tissues. Moreover, using different infill densities, the printed materials were able to achieve a similar radiodensity range as obtained in different tissue-equivalent inserts in the Gammex phantom (238 HU to -673 HU). The reproducibility results showed good agreement between the HU values of the replicas compared to the original test phantoms, confirming the duplicability of the printed materials.

1. Introduction

Additive manufacturing (AM), also known as three-dimensional (3D)-printing, is increasingly used to fabricate complex structures from physical models generated from three-dimensional (3D) computer-aided design (CAD) data [1]. Recently, 3D-printing has also become popular for the development of medical imaging phantoms [2-4]. The material used in such imaging phantoms should ideally mimic physical and imaging characteristics as close as possible to human tissue [5, 6]. The radiological properties of any tissue equivalent material can be characterized by Hounsfield Unit (HU) which represents linear attenuation coefficient of X-rays in CT [7]. Solid materials such as Polymethyl methacrylate (PMMA) and resins [8, 9, 10] have been proposed to mimic soft tissues HUs for a breathing thorax and pelvis CT phantoms. In addition, several rigid 3D printable materials have been examined regarding to their radiological properties e.g. Acrylonitrile butadiene styrene (ABS), Polylactic acid (PLA), Polyamid 12 (Nylon12 or PA12), Acrylonitrile styrene acrylate (ASA Pro), Polyethylenterephthalat (PETG), Vero PureWhite and VeroClear [11-14]. However, flexible soft tissue materials are needed for imaging phantoms used for different scenarios where anatomical deformations should be simulated, for example motion-adaptive radiotherapy treatments [15, 16], phantoms with respiratory motion for needle-based liver interventions [17] and phantoms for different needle-based surgery training [18, 19, 20].

Several studies have proposed deformable imaging phantoms. Various types of non-printable flexible materials were investigated with the goal of soft-tissue phantom constructions, e.g. gelatin, silicone and urethane materials. Gelatin materials have shown similar radiological properties to soft tissue [21, 22]. However, the drawback of such gelatin-based phantoms is that the radiological properties usually change over time, as they are losing water. Synthetic polymers, for example polyvinyl chloride (PVC) and silicone have generally more stable properties and also a longer shelf life because they do not have water within the structure [23]. It was also shown that PVC with different softener ratios can result in different HU, allowing the replication of many organ densities [22, 24, 25]. Furthermore, silicone and urethane materials have been used to study soft tissues, with different types and mixtures providing the ability to adjust concentrations to mimic the relevant HU [15, 22, 26, 27]. Although the aforementioned flexible materials [21-27] showed similar radiological properties to soft tissues and organs, they are only used to fill phantoms developed using molding techniques, so direct 3D printing of deformable phantoms was not possible.

In medical imaging, 3D printing technology is now being explored for directly creating phantoms from flexible, 3D-printable tissue-like materials. 3D printable rubber-elastomeric polymer called PORO-LAY filaments were used to develop tissue mimicking materials for MRI phantoms [28]. A 3D-FDM printer with dual extruders was used for printing. In another study [29] radiological characteristics of PORO-LAY filament materials were investigated by measuring their HUs values at different infill structures, infill densities and introduction of several kinds of fluids.

Different silicone additive manufacturing technologies have already used to manufacture anatomical models based on extrusion techniques [30-34]. In [31] authors proposed an additive manufacturing technique for fabrication of tissue-like aortic heart valves with customized geometry. They used biocompatible silicones with tunable mechanical properties and fabricated heart valves by combining spray and extrusion-based additive manufacturing processes. Luis et al [32, 33] used a heat-cured extrusion-based method using a custom-made 3D silicone printer device and two-part Ecoflex silicone resins in order to 3D-print standard-shaped silicone coupon and irregular-shaped meniscus structures as well silicone meniscal implants. Yin et al [34] developed an extrusion-based 3D printing system that consists of three printing nozzles where rigid and elastomeric materials were combined. They designed phantoms consisting of multiple layers, where Acrylonitrile Butadiene Styrene (ABS) filament, Thermoplastic Polyurethane (TPU) filament and silicone ink were selected as printing materials for bone, soft tissue and skin, respectively. Recently, we developed a multi-material extrusion 3D-printer which is capable of printing a 1k-silicone and an additional viscous fluid, combined with standard filament printhead. Various objects and anatomical models were printed successfully using this combined 3D printed technology [35-36]. In the above studies [30-36], radiological properties of the 3D printed silicone materials were not investigated.

The use of 3D-printed silicone materials in imaging phantoms can be very beneficial for scenarios where flexibility and softness of materials is important [15-19, 29, 31-34]. To our knowledge, is no systematic study in the literature investigated the radiological properties of silicone materials/fluids produced directly by extrusion 3D printing for imaging phantoms. In addition, the influence of different infill densities (ratio of silicone to air) on the resulting radiodensity in CT imaging using different filament materials has been extensively studied [12, 29, 37], but no such study has been reported for printable silicon materials. The aim of this study is to evaluate the radiological properties of extrusion silicone 3D printer phantoms in CT imaging; The HU values of 3D-printed test phantoms consisting of three different silicone printing materials are evaluated by varying the infill density to modify their radiological properties and a gradient of radiodensity that mimics different soft tissues and organs is introduced. A comparison with the Gammex Tissue Characterization Phantom is provided to compare the resulting HU values of the 3D-printed test silicone phantoms with those of a standard phantom. Furthermore, additional experiments using an anatomical phantom consisting of different infill structure densities mimicking realistic human CT radiodensity at different organs is performed to compare the resulting HU with the corresponding soft tissues in the real CT scan. A reproducibility analysis to ensure the reproducibility of the radiation density of the proposed phantoms is also performed.

## 2. Materials and Methods

### 2.1. Silicone 3D printing

In the current study, a custom-built silicone 3D-printer, based on a commercially available Railcore II 300 ZL filament printer was used (Fig.1). The moving carriage of this base printer was extended with a Viscotec Vipro-HEAD 3/3 fluid printhead, which consists of two independently controllable extruders. This way, the printhead can deposit either two high-viscosity fluids independently, or alternatively a two-component material. Further details about this 3D-printer are found in [35, 36]. In this study, three single-component (1k) condensation-crosslinking liquid silicone rubbers (supplied by Elkem Silicones SAS, Lyon, France) were used with 0.41 mm nozzles on the fluid extruders. The individual print runs were planned in the open-source slicing software PrusaSlicer v2.3.0. The silicone printhead was calibrated according to calibration method described in [36].

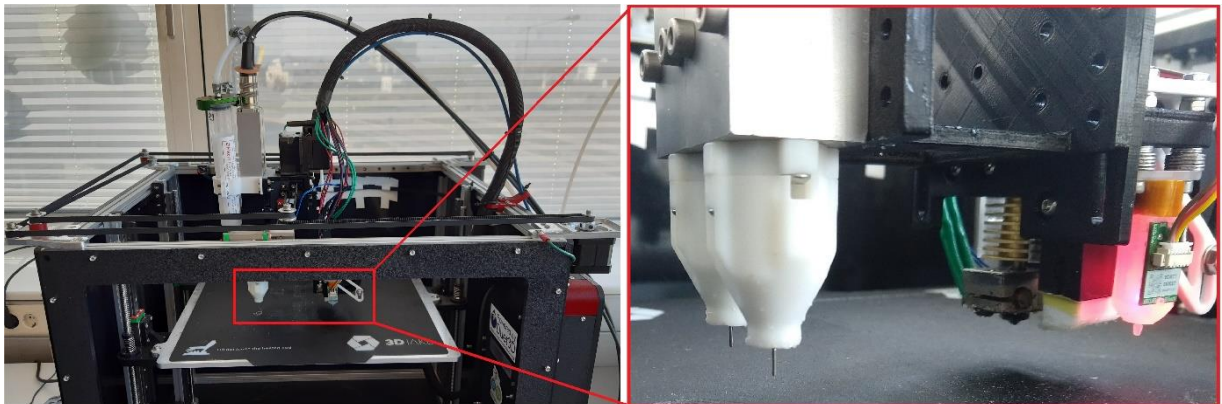

Figure 1: (Left image) the multi-material printer used in this study. (Right image) both independently controllable fluid extruders are shown.

### 2.2. Test phantoms

To explore the connection between infill structure density and radiological properties, several rectangular blocks (called test phantoms in this study) of 14 x 14 x 20 mm were printed from the three silicone rubber materials, namely Elkem AMSil 20101 (Material 1), 20102 (Material 2) and 20103 (Material 3). The various infill structure densities covered a range of 30% to 100% material volume fraction, with 10% increments between 30% and 60%, 5% increments between 65% and 75%, and 3% increments between 79 and 100% yielding 15 levels in total. The corresponding test phantoms are called S1-S15 accordingly, with S1 belonging to 100% infill and S15 to 30% infill (Fig. 2). The 100% infill test phantom was printed with a rectilinear infill structure, while all the other test phantoms were printed with a gyroid infill structure generated in PrusaSlicer software (Fig. 2). The levels of 40%, 70% and 91% were printed 6 times of each material to evaluate reproducibility, while all other levels were printed once of each material. All blocks were printed with a 20 mm/s printing speed, two solid closing layers on top and bottom, two contour lines on the sides, in ambient conditions of 20-30 °C and 65-85% relative humidity. The blocks were removed from the building platform 24 hours after printing to allow sufficient crosslinking. As a quality control measure, the blocks were weighed on a KERN PES 42002M laboratory scale. The weight was compared to the estimations of PrusaSlicer to ensure that all printed specimens fall within a relative weight error range of ±5%.

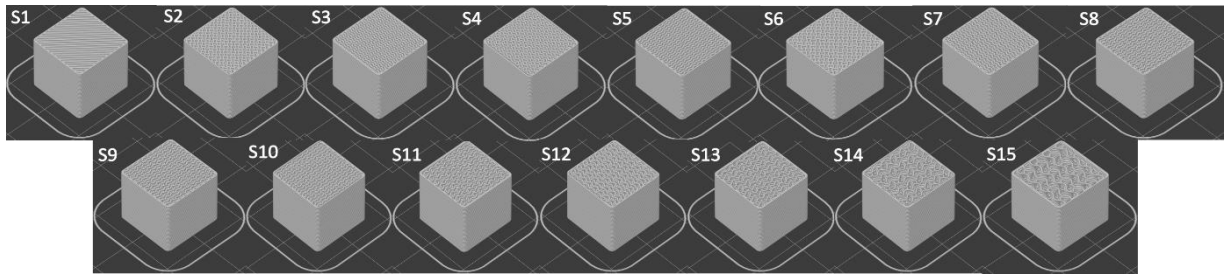

Figure 2. The gyroid infill structures generated in PrusaSlicer software for different infills including S1-S15 corresponding to 100% to 30% material volume fraction, with 10% increments between 30% and 60%, 5% increments between 65% and 75%, and 3% increments between 79 and 100%.

## 2.3. Anatomical use case

As a first use case of the HU/infill density mapping acquired through examining the test phantoms, a radiological phantom representing a simplified scaled down anatomical model derived from an abdominal CT [11] was printed using the Elkem AMSil 20101 (translucent) and 20102 (white) silicones. The abdominal CT was processed using 3D Slicer software 4.11.2 (Boston, MA, USA). Major anatomic structures were roughly segmented using the region-growing algorithm, distinguishing the bone, kidney and vessels, liver and spleen, connective tissue, muscle and skin as well as the air in lungs and abdomen (Fig. 3). The approximate HU values within these structures were 759, 175, 96, 32, −65, and −794, respectively. One slice of the torso was selected from this dataset and within this slice, segmentation was manually refined. The segments were smoothened and exported in STL format.

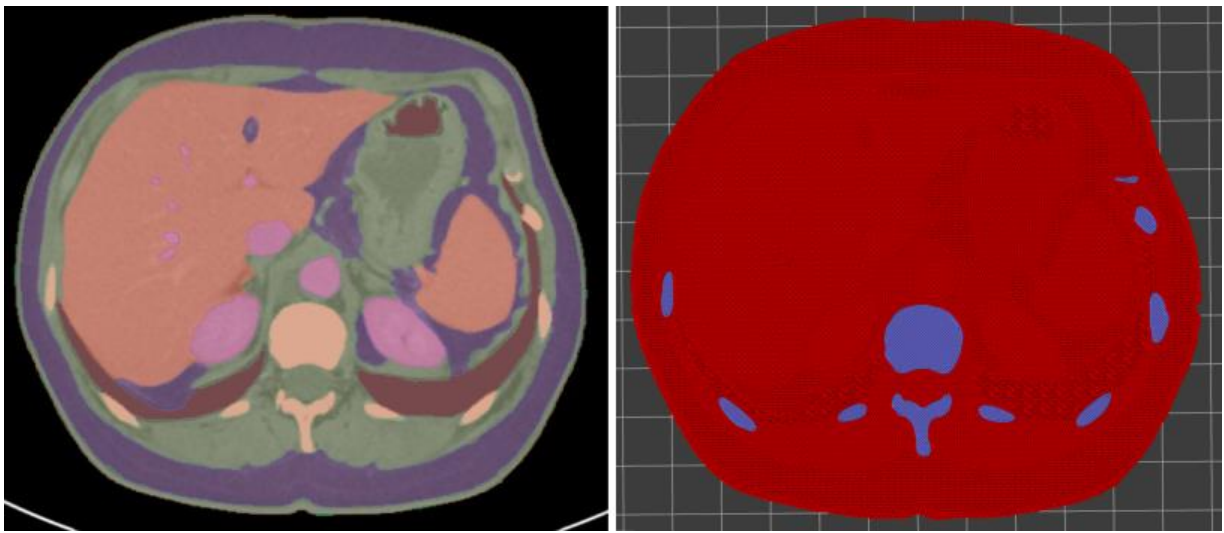

Figure 3: Left: Segmented abdominal slice, separating the bone (yellow), kidney and vessels (magenta), liver and spleen (red), connective tissue (blue), muscle, skin and abdomen (green) as well as the air in the lungs and abdomen (grey). Right: The segments were reassembled in Prusa Slicer and their corresponding gyroid infill percent was assigned to them.

The exported segments were post-processed in Autodesk Meshmixer v3.5 to correct smaller meshing errors. Afterwards, they were reassembled in PrusaSlicer (Fig. 3). According to the approximate HU values determined for each tissue, the corresponding air/material ratios were introduced as a gyroid infill volume fraction to each segment, based on the prior radiodensity analysis of the test phantoms (described in 2.2). The resulting radiological phantom slice was printed with a 45% layer-plane scaling and a 7.5 mm thickness. Both halves of the fluid printhead were used, one loaded with AMSil 20102 to print the bone segments and the other with AMSil 20101 to print all other segments with their corresponding gyroid infill densities.

## 2.4. HU analysis

All the test phantoms as well as the anatomical phantom were scanned with the standard clinical CT protocol (SOMATOM Definition AS, Siemens Healthineers, Erlangen Germany, tube current time product 315 mAs, tube voltage 120 kVp, slice thickness 0.60 mm, pixel spacing 0.77 mm) and the resulted radiodensities were computed. Additional scans at 100 kVp and 80 kVp were also performed from all scans in order to report the resulting HU at different energy spectra. In case of the anatomical phantom the obtained HU values were compared with those of the original human CT data. Analyze 12.0 toolkit (AnalyzeDirect, Over-land Park, United States) was used in order to measure the HU value from the test phantoms and anatomical phantom CT scans. Different line profiles inside some regions of interests within the phantom images were selected and the HU values were computed by calculating the average and the standard deviation over all points for the selections related to those line profiles. In addition, for each test phantom. Signal-to-noise-ratio (SNR) was also calculated by dividing the achieved average of HU values to the standard deviation of HU values, the absolute values were reported.

## 2.5. Comparison with standard Gammex tissue equivalent phantom

The resulting HU values from the developed 3D printed phantoms were compared with a standard phantom, the Gammex Tissue Characterization Phantom (Gammex Model 467, Middleton, USA). A CT scan from the standard Gammex phantom was obtained with the same CT parameters (only at 120 kVp) as for 3D printed test phantoms (Section 2.4). The average and standard deviation (SD) of the resulting HU related to the Gammex tissue equivalent inserted cylinders including bone mineral, inner bone, liver, brain, solid water, breast, adipose and lung were calculated using Analyze 12.0 toolkit with the same method as in Section 2.4. The relation between different tissue equivalent inserted cylinders and the printed test phantoms with different ratios was also investigated.

## 2.6. Mass density measurement

The mass density of all 3D printed test phantoms was calculated by dividing the weight (g) of each printed test phantom by the volume (cm3) of that phantom. The weight in g was measured on a KERN EMB 200-3 laboratory scale (Kern&Sohn GmbH, Balingen, Germany). The volume of each test phantom was calculated from the CT scan of each phantom; first a thresholding step was applied in 3D-Slicer and each phantom was segmented and cropped, accordingly. Afterward, volume of each cube in cm3 was calculated using the Label Statistics module in 3D-Slicer. The reported density for Material 1, Material 2 and Material 3 according to their technical data sheets are 1.01 g/ cm3, 1.30 g/ cm3 and 1.04 g/cm3, respectively.

## 2.7. Reproducibility of the silicone test phantoms

A reproducibility analysis was also performed to ensure the reproducibility of the radiation density of the proposed phantoms. For this aim, the infill density levels of 40%, 70% and 91% were printed 6 times (replicas) of each material to evaluate reproducibility. The replicas were scanned using the same CT parameters as for the 3D printed test phantoms (Section 2.4) and also HU values (in terms of average and standard deviation) were calculated with the same HU analysis as described in Section 2.4.

## 3. Results
### 3.1. 3D printed test phantoms results

The three materials with different infill densities S1-S15 (Section 2.2) were printed successfully (Fig. 4). The CT scan from the test phantoms related to the three materials was acquired (Fig. 5). The corresponding HUs related to scan with 120 kVp were calculated (Table 1). According to the results from Table 1, a range between -639 to 252 HUs, -485 to 770 HUs and -612 to 316 was achieved for Material 1, Material 2 and Material 3, respectively. Lower HU values were achieved using Material 1 and Material 3 compared to Material 2. The resulting HU values for the other two different scan settings including 100 kVp and 80 kVp were also obtained (Fig. 6). For all materials and infill settings a decrease in HU with increasing X-ray energy was observed.

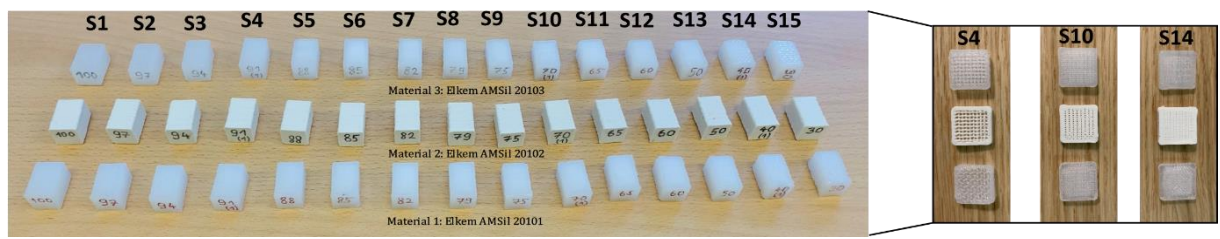

Figure 4: The 3D printed test phantoms related to the three printed materials (down: Material 1, middle: Material 2, up: Material 3) at different infill densities (S1-S15) and the corresponding cross sections at S4, S10 and S14.

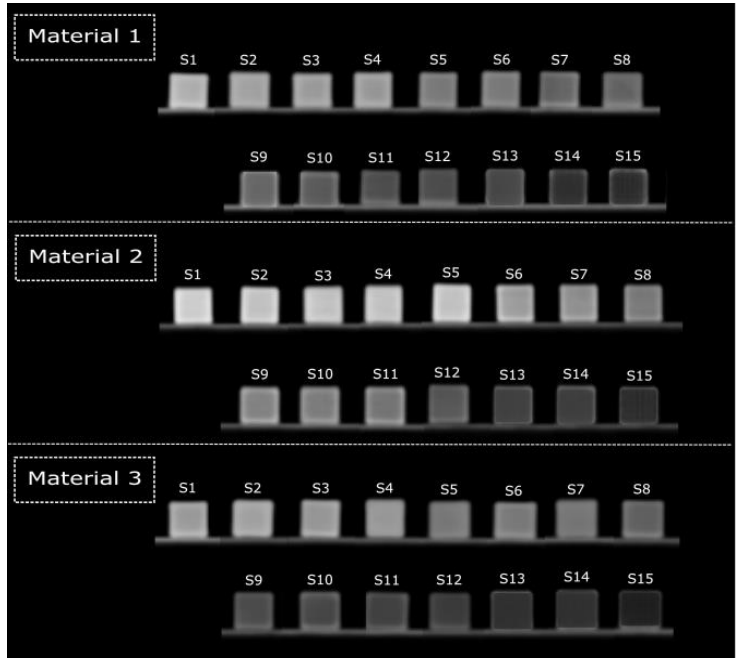

Figure 5: The axial view of the CT scan from all test phantoms related to Material1, Material 2 and Material 3 at different infill densities S1-S15.

**Table 1:** Hounsfield Unit and standard deviation and signal-to-noise ratio related all three materials at different infill densities at 120 kV. Mat.1: Material 1, Mat.2: Material 2, Mat.3: Material 3.

| Infill density (%) | Hounsfield Unit (HU) | | | Standard deviation (SD) | | | Signal-to-noise-ratio (SNR) | | |
|---|---|---|---|---|---|---|---|---|---|
| | Mat.1 | Mat.2 | Mat.3 | Mat.1 | Mat.2 | Mat.3 | Mat.1 | Mat.2 | Mat.3 |
| **100 (S1)** | 252.45 | 770.85 | 316.36 | 1.87 | 2.66 | 1.05 | 135 | 289.79 | 301.29 |
| **97 (S2)** | 196.20 | 699.19 | 313.07 | 2.47 | 2.77 | 1.13 | 79.43 | 252.41 | 277.05 |
| **94 (S3)** | 158.11 | 591.41 | 290.81 | 2.61 | 2.74 | 2.03 | 60.57 | 215.84 | 143.25 |
| **91 (S4)** | 158.43 | 586.34 | 212.33 | 2.91 | 3.33 | 2.26 | 54.44 | 176.07 | 93.95 |
| **88 (S5)** | 142.19 | 603.84 | 201.70 | 3.02 | 3.24 | 3.65 | 44,43 | 186.37 | 55.26 |
| **85 (S6)** | 45.44 | 500.13 | 70.21 | 3.30 | 3.20 | 2.20 | 13.76 | 227.33 | 31.91 |
| **82 (S7)** | 26.62 | 505.39 | 58.92 | 1.86 | 3.95 | 2.64 | 14.31 | 127.94 | 22.32 |
| **79 (S8)** | 11.92 | 441.05 | 30.24 | 2.11 | 3.17 | 2.19 | 2.65 | 139.13 | 13.80 |
| **75 (S9)** | -40.08 | 357.41 | 19.47 | 2.21 | 2.54 | 2.01 | 18.14 | 140.71 | 9.68 |
| **70 (S10)** | -153.05 | 202.12 | -95.37 | 1.82 | 1.95 | 1.29 | 84.09 | 103.65 | 73.93 |
| **65 (S11)** | -212.81 | 68.07 | -206.87 | 1.52 | 1.86 | 1.50 | 140.00 | 36.58 | 137.91 |
| **60 (S12)** | -223.35 | 00.31 | -198.06 | 1.49 | 2.05 | 1.46 | 149.89 | 00.15 | 135.65 |
| **50 (S13)** | -354.20 | -165.87 | -353.88 | 1.66 | 2.14 | 1.65 | 213.37 | 77.50 | 214.47 |
| **40 (S14)** | -498.96 | -298.10 | -496.14 | 1.68 | 2.13 | 1.68 | 297.00 | 139.95 | 295.32 |
| **30 (S15)** | -639.19 | -485.18 | -612.55 | 2.06 | 2.56 | 2.16 | 310.28 | 189.51 | 283.58 |

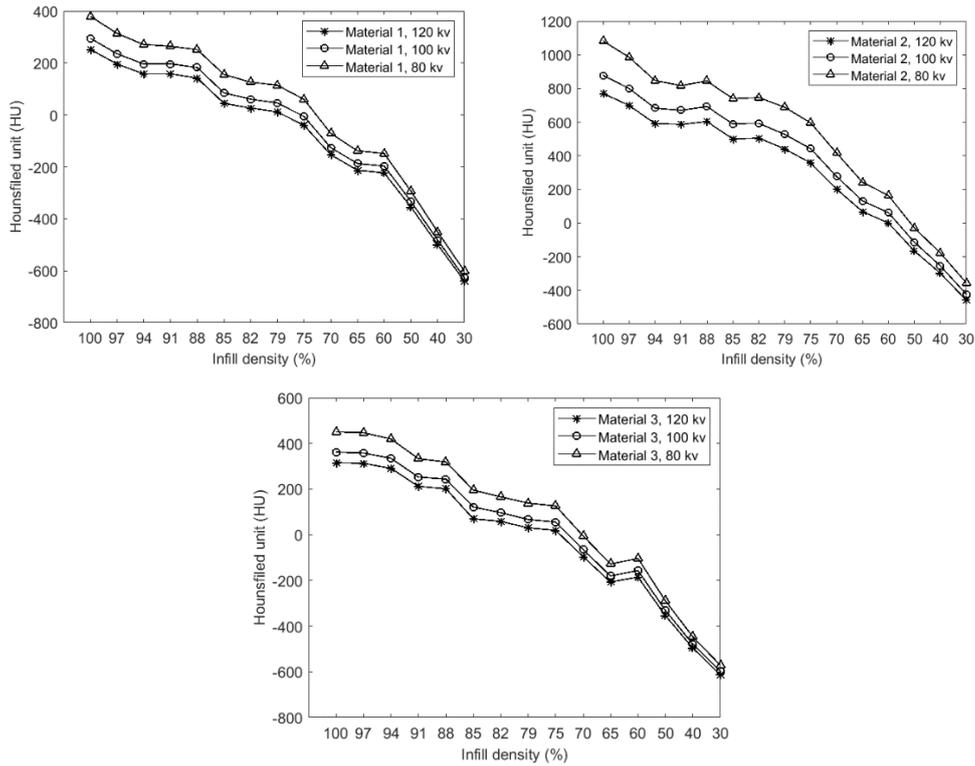

Figure 6: Resulting HU values for three materials for the three different scan settings including 120 kVp, 100 kVp and 80 kVp.

### 3.2. Comparison of the resulting HU with the Gammex phantom

Axial slice of the CT scan of the Gammex phantom is shown in Fig. 7. HU values for different tissue equivalent inserts inside the Gammex phantom including bone mineral, inner bone, liver, brain, solid water, breast, adipose and lung were calculated using the same method as described in Section 2.4 (Table 2). For all three printed materials, the closest achieved HU to the Gammex phantom inserts and the corresponding infill density is reported in Table 3. According to the results, the test phantoms could achieve similar radiodensity range as achieved in the Gammex phantom including 238 HU to -673 HU (bone mineral to lung) when using different infill densities.

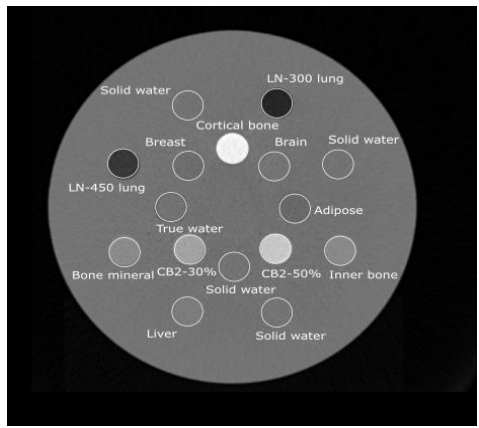

Figure 7: Axial slice of the CT scan of the Gammex phantom including different tissue equivalent inserts.

**Table 2:** Table showing the mean Hounsfield Unit (HU), standard deviation (SD) and signal-to-noise ratio (SNR) of the resulting HU related to the Gammex tissue equivalent inserted cylinders including bone mineral, inner bone, liver, brain, solid water, breast, adipose and lung and the corresponding materials and infill densities with the closest HUs from the printed test phantoms. Mat.1: Material 1, Mat.2: Material 2, Mat.3: Material 3.

| Gammex phantom | HU+SD, SNR | Printed Material (Infill density (%)) | HU+SD, SNR |
|---|---|---|---|
| **Bone mineral** | 238.24 ± 17, 14.01 | Mat.1 (100) | 252.45± 1.87,135 |
| | | Mat.2 (70) | 202.12± 1.95, 103.65 |
| | | Mat.3 (91) | 212.33± 2.26, 93.95 |
| **Inner bone** | 210.56 ± 12, 17.55 | Mat.1 (97) | 196.20± 2.47,79.43 |
| | | Mat.2 (70) | 202.12± 1.95, 156.68 |
| | | Mat.3 (88) | 201.70± 3.65, 55.26 |
| **Liver** | 66.14 ± 10, 6.64 | Mat.1 (85) | 45.44± 3.30, 13.76 |
| | | Mat.2 (65) | 68.07±1.86, 36.58 |
| | | Mat.3 (82) | 58.92± 2.64,22.32 |
| **Brain** | 24.72 ± 6, 4.12 | Mat.1 (82) | 26.62± 1.86, 14.31 |
| | | Mat.2 (60) | 00.31± 2.05, 00.15 |
| | | Mat.3 (79) | 30.24± 2.19,13.80 |
| **Solid Water** | -1.38 ± 3, 0.46 | Mat.1 (79) | 11.92± 2.11,2.65 |
| | | Mat.2 (60) | 00.31± 2.05,00.15 |
| | | Mat.3 (75) | 19.47± 2.01,9.68 |
| **Breast** | -72.47 ± 10, 7.24 | Mat.1 (75) | -40.08± 2.21, 18.14 |
| | | Mat.2 (60) | 00.31± 2.05,00.15 |
| | | Mat.3 (70) | -95.37± 1.29, 72.37 |
| **Adipose** | -109.30 ± 13, 8.40 | Mat.1 (70) | -153.05± 1.82, 84.09 |
| | | Mat.2 (50) | -165.87± 2.14, 77.50 |
| | | Mat.3 (70) | -95.37± 1.29, 72.37 |
| **Lung-450** | -475.64 ± 12, 39.63 | Mat.1 (50) | -498.96± 1.68, 297.00 |
| | | Mat.2 (30) | -485.18± 2.56, 189.51 |
| | | Mat.3 (40) | -496.14± 1.68, 295.32 |
| **Lung-300** | -673.42 ± 15, 44.89 | Mat.1 (30) | -639.19± 2.06, 310.28 |
| | | Mat.2 (30) | -485.18± 2.56, 189.51 |
| | | Mat.3 (30) | -612.55± 2.16, 283.58 |

### 3.3. 3D printed anatomical phantom results

The anatomical phantom was printed successfully (Fig. 8A). According to the HU values achieved in Section 3.1, we assigned the materials and the infill densities which corresponded to the organs/tissues radiodensity for the anatomical phantom (Table 3). The axial slice of the resulting CT scan from the anatomical phantom is shown in Fig. 8B and compared with patient CT (Fig. 8C). The resulting HU values were calculated (same method as in Section 2.4) from the CT scan from this phantom (Table 3).

According to the results of Table 3, a good agreement between the target HU values (approximate HU values from abdomen CT) and the fabricated phantom HU values was achieved at all tissues within the phantom.

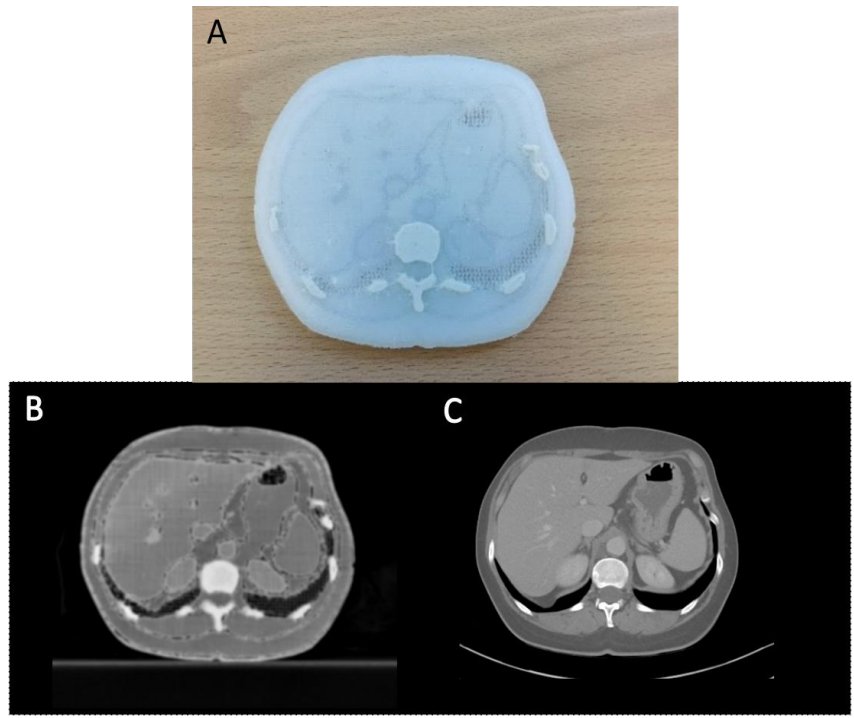

Figure 8: A) the 3D printed anatomical phantom, B) axial slice of the CT scan from the 3D printed anatomical phantom, C) axial slice of the abdominal CT scan.

**Table 3:** The mean and standard deviation (SD) of the resulting HU of the anatomical phantom compared to target HU values (approximate HU from abdomen CT) as well as corresponding test phantom materials and infill densities used for different tissues.

| Tissue type | Material, infill density | Patient HU | Anatomical phantom HU |
|---|---|---|---|
| **Bone** | Mat 2, 100% | 759.23±54.12 | 727.26±3.27 |
| **Kidney and vessels** | Mat 1, 97% | 175.36±34.31 | 151.46±2.78 |
| **Liver and spleen** | Mat 1, 88% | 96.12±30.56 | 102.51±4.28 |
| **Muscle** | Mat 1, 82 % | 32.84±27.52 | -17.24±3.35 |
| **Connctive tissue** | Mat 1, 75 % | -65.74±38.46 | -92.18±4.15 |
| **Lung** | Mat 1, 30% | -794.52±29.84 | -680.28±3.18 |

### 3.4. Resulting mass densities of test phantoms

The calculated density values (g/cm3) of all 3D printed test phantoms are also presented in Table 4. As can be seen from the results of Table 4, the test phantoms with the infill densities that contained a higher percentage of air had lower density values, which was due to a lower weight.

**Table 4.** Density values (g/cm3) for all 12 tumor phantoms.

| Infill density (%) | Material 1 (g/cm3) | Material 2 (g/cm3) | Material 3 (g/cm3) |
|---|---|---|---|
| 100% (S1) | 1.003 | 1.225 | 0.997 |
| 97% (S2) | 0.956 | 1.167 | 0.941 |
| 94% (S3) | 0.910 | 1.140 | 0.937 |
| 91% (S4) | 0.906 | 1.102 | 0.923 |
| 88% (S5) | 0.854 | 1.071 | 0.897 |
| 85% (S6) | 0.821 | 1.067 | 0.850 |
| 82% (S7) | 0.819 | 1.045 | 0.827 |
| 79% (S8) | 0.783 | 1.003 | 0.799 |
| 75% (S9) | 0.760 | 0.984 | 0.769 |
| 70% (S10) | 0.743 | 0.963 | 0.728 |
| 65% (S11) | 0.691 | 0.904 | 0.705 |
| 60% (S12) | 0.654 | 0.854 | 0.683 |
| 50% (S13) | 0.569 | 0.731 | 0.600 |
| 40% (S14) | 0.477 | 0.713 | 0.500 |
| 30% (S15) | 0.427 | 0.618 | 0.442 |

### 3.5. Resulting radiodensity for the replicas

The average HU and standard deviation over the CT scan from the 6 replicas related to the three materials at 40%, 70% and 91% infill densities were calculated (Table 5). The mean HU values found over the replicas for all materials and infill densities revealed a good reproduction of the original densities (Table 1).

**Table 5:** The average HU and standard deviation over the CT scan from the 6 replicas related to the three materials at 40%, 70% and 91% infill densities.

|  | Infill density= 91 % | | | Infill density= 70 % | | | Infill density= 40 % | | |
|---|---|---|---|---|---|---|---|---|---|
|  | Mat.1 (HU) | Mat.2 (HU) | Mat.3 (HU) | Mat.1 (HU) | Mat.2 (HU) | Mat.3 (HU) | Mat.1 (HU) | Mat.2 (HU) | Mat.3 (HU) |
| **Replica 1** | 158.43 | 586.34 | 212.33 | -153.05 | 202.12 | -95.37 | -498.96 | -298.10 | -496.14 |
| **Replica 2** | 111.12 | 575.35 | 183.88 | -127.09 | 228.23 | -66.98 | -502.02 | -319.37 | -498.32 |
| **Replica 3** | 158.28 | 539.20 | 183.85 | -139.11 | 188.69 | -84.89 | -492.62 | -317.87 | -527.75 |
| **Replica 4** | 111.65 | 524.05 | 256.35 | -166.41 | 236.75 | -115.88 | -494.62 | -278.58 | -490.98 |
| **Replica 5** | 162.36 | 558.91 | 168.97 | -107.63 | 219.22 | -101.18 | -497.24 | -290.86 | -479.42 |
| **Replica 6** | 159.06 | 610.73 | 131.09 | -174.72 | 225.92 | -81.97 | -490.66 | -254.32 | -476.14 |
| **Mean HU** | 143.48 | 565.76 | 189.40 | -144.56 | 216.82 | -91.05 | -496.26 | -293.18 | -494.79 |
| **Standard deviation HU** | 24.90 | 31.71 | 42.18 | 25.31 | 18.01 | 16.95 | 4.20 | 24.69 | 18.43 |

## 4. Discussion and conclusion

3D-printing technology is widely used in producing medical imaging phantoms. Anthropomorphic 3D printed imaging phantoms mimicking tissues and contrasts in real patients with regard to X-ray attenuation can be very valuable to produce phantoms for image quality optimization, comparison of image quality between different imaging systems, dosimetry, quality control and imaging protocol definition. In this study, the radiological properties of extrusion silicone 3D printer phantoms in CT imaging were investigated. The HU values of three different silicone printing materials at changing the infill density were assessed in order to modify their radiological properties and to achieve a spectrum of radiodensity that mimics different soft tissues and organs. The 3D printed test phantoms made of three different materials were able to produce a HU spectrum in the range of -639 to +780 HU in CT at a scan setting of 120 kVp, which corresponds not only to the radiation attenuation of human soft tissue, but also of bones and lungs. We also investigated the performance of our approach using an anatomical phantom (generated from a human abdominal CT) containing complex structures and combinations of different infill density structures and materials within the same phantom. We observed good agreement between the HU target values in the abdominal CT and the HU of the 3D printed anatomical phantom in all tissues.

So far, several phantoms mimicking soft tissue radiodensities in CT imaging have been reported in the literature and their radiological properties were evaluated; however, these phantoms were mainly made of rigid materials [11-14]. Deformable 3D-printed imaging phantoms can be very beneficial for various scenarios [15-20]. Recently, different silicone additive manufacturing technologies have already been proposed to manufacture flexible and soft anatomical models based on extrusion techniques [30-36]. To our knowledge, there is no systematic study reported in the literature that explores the radiological properties of 3D direct printing of silicone materials/fluids for imaging phantoms manufactured using extrusion-based printer. Furthermore, the impact of different infill densities to adjust radiodensity in CT imaging on such printable silicon materials has not been investigated. Our study investigated the radiological properties of the 3D printed silicone phantoms directly printed based on extrusion technique. A good similarity was also observed between the radiodensity of the silicone test phantoms and different tissue-equivalent Gammex phantom inserts. In total, lower standard deviation and higher absolute value of SNR was observed in the test phantoms compared to the different tissue equivalent inserts inside the Gammex phantom. This suggests the efficiency of the 3D printed silicone materials in producing CT imaging phantoms. Reproducibility test results also showed a good agreement between the HU values of the replicas compared to the initial test phantoms, confirming duplicability of the printed materials.

By changing the beam energies, an inverse relationship between the attenuation coefficient and beam hardness was observed for all samples which is typical for bone tissue but is in contrast to soft tissues which exhibit constant HU over different beam energies [39]. This indicates a higher effective atomic number of the phantoms compared to water. Future developments will investigate other types of flexible printable materials/liquids which mimic the attenuation of tissues over a large X-ray energy range. Introduction of fluid in the printed structure may also be used to adjust for different radiodensity values. This could be done by either injecting fluid into the infill structure layer-wise during printing or injecting it into the whole phantom manually after printing. Such a strategy is considered as the future perspective of this study. In addition, investigating of the mechanical properties of test phantoms with different infill densities is also a promising future direction of this research. Certain infill structures may provide both matching HUs and matching mechanical properties at the same time for certain tissues. Such cases may be used to create radiologically and mechanically realistic models, which may be especially useful in surgical training or image guided procedures where flexible phantoms are needed.

**Acknowledgements**

This work has been supported by ACMIT – Austrian Center for Medical Innovation and Technology, which is funded within the scope of the COMET program and funded by Austrian BMVIT and BMWFW and the governments of Lower Austria and Tyrol. This work was also supported by the Provincial Government of Lower Austria (Land Niederösterreich) under grant assignment number WST3-F2- 528983/005-2018.